\documentclass[
    onecolumn,
    preprintnumbers,
    amsmath,
    amssymb,
    prd,
    showpacs
]{revtex4}
\usepackage{graphicx}%
\usepackage{amsmath}
\usepackage{amsthm}
\usepackage[normalem,normalbf]{ulem}
\usepackage{amssymb}
\usepackage{mathrsfs}
\usepackage{bm}
\usepackage{enumerate}
\newcommand{\del}{\partial}
\newcommand{\nn}{\nonumber}

\newcommand{\wt}{\widetilde}
\newcommand{\half}{\frac12}
\newcommand{\halfi}{\frac{i}{2}}
\newcommand{\beq}{\begin{equation}}
\newcommand{\eeq}{\end{equation}}
\newcommand{\beqa}{\begin{eqnarray}}
\newcommand{\eeqa}{\end{eqnarray}}
\newcommand{\bseq}{\begin{subequations}}
\newcommand{\eseq}{\end{subequations}}

\newcommand{\fr}{\frac}
\newcommand{\mn}{{\mu \nu}}
\newcommand{\thetak}{\theta^{\alpha_1 \beta_1}\cdots
\theta^{\alpha_k \beta_k}}
\newcommand{\ppa}{[P_{\alpha_1},[P_{\alpha_2},
\cdots,[P_{\alpha_k},}
\newcommand{\ppb}{[P_{\beta_1},[P_{\beta_2},
\cdots,[P_{\beta_k},}
\newcommand{\bra}[1]{\langle #1|}
\newcommand{\ket}[1]{|#1\rangle}
\newcommand{\braket}[2]{\langle #1|#2\rangle}
\newcommand{\phiqn}{\ket{q_1,\cdots, q_n}}

\newcommand{\psiqn}{\ket{q_1,\cdots, q_n}_\star}

\newcommand{\ad}{a^\dagger}
\newcommand{\cA}{\mathcal{A} }
\newcommand{\cE}{\mathcal{E}}
\newcommand{\cF}{\mathcal{F}}
\newcommand{\cH}{\mathcal{H}}
\newcommand{\cP}{\mathcal{P} }
\newcommand{\cT}{\mathcal{T} }
\newcommand{\cU}{\mathcal{U}}
\newcommand{\cUP}{\mathcal{U} (\mathcal{P} )}
\newcommand{\cUFP}{\mathcal{U}_{\mathcal{F}} (\mathcal{P})}
\newcommand{\mcP}{\mathscr{P}}
\newcommand{\btr}{\blacktriangleright}
\newcommand{\phase}{\cE(q_1,\cdots,q_n)}

\begin{document}
%-------------------------%-------------------------------
\title{Noncommutative Field Theory from twisted Fock space }
%---------------------------------------------------------
\author{Jong-Geon Bu}
\email{bjgeon@yonsei.ac.kr}
%--------------------------------------------------------
\author{Hyeong-Chan Kim}
\email{hckim@phya.yonsei.ac.kr}
%---------------------------------------------------------
\author{Youngone Lee}
\email{youngone@phya.yonsei.ac.kr}
%--------------------------------------------------------
\author{Chang Hyon Vac}
\email{Shoutpeace@yonsei.ac.kr}
%----------------------------------------------------
\author{Jae Hyung Yee}%
\email{jhyee@phya.yonsei.ac.kr}
%---------------------------------------------------
\affiliation{Department of Physics, Yonsei University, Seoul, Korea.
}%
%----------------------------------------------------
\date{\today}%
%---------------------------------------------------
\bigskip
%----------------------------------------------------
\begin{abstract}
%-----------------------------------------------------
\bigskip

We construct a quantum field theory in noncommutative
spacetime by twisting the algebra of quantum operators
(especially, creation and annihilation operators)
of the corresponding quantum field theory in commutative
spacetime.
The twisted Fock space and S-matrix consistent
with this algebra have been constructed. The resultant
S-matrix is consistent with that of Filk\cite{Filk}.
We find from this formulation that the
spin-statistics relation is not violated
in the canonical noncommutative field theories.
%-------------------------------------------------
\end{abstract}
%-------------------------------------------------
%------------------------------------------------
%------------------------------------------------
\maketitle
%-------------------------------------------------

%%%%%%%%%%%%%%%%%%%%%%%%%%%%%%%%%%%%%%%%%%
\section{Introduction}
%%%%%%%%%%%%%%%%%%%%%%%%%%%%%%%%%%%%%%%%%%%%
There have been much interests in field theories
\cite{Behr, Bayen, Douglas, Doplicher}
in noncommutative spacetime
where coordinates satisfy
the commutation relation,
\beq
[x^\mu, x^\nu]= i \theta^{\mu\nu},
\label{nc}
\eeq
with $\theta^{\mu\nu}$ being an anti-symmetric constant.
By the Weyl-Moyal correspondence \cite{Weyl},
a field theory in noncommutative spacetime
satisfying Eq.~(\ref{nc})
is transformed to a field theory
in commutative spacetime
in which product of two spacetime functions
is defined by the Moyal product,
\beq
\label{Moyal}
(\phi * \psi) (x) =\left.
e^{\halfi \theta^{\mu\nu}
\partial_{x^\mu}\partial_{y^\nu}}
\phi (x)\cdot \psi (y)
\right|_{x=y}.
\eeq
Due to the existence of noncommutativity parameter
the theory is not Poincar\'{e} invariant.
The properties such as causality and unitarity
are suggested to be violated in the presence of
spacetime noncommutativity (STNC)
\cite{Seiberg},\cite{Gaume},\cite{chaichian0},\cite{Gomis},\cite{Bahns},
while they remain satisfied in
the space-space noncommutative(SSNC) case.
There have been some attempts to cure these problems
%Time ordered perturbation theory
\cite{Bahns1},\cite{Bahns2},\cite{BahnDo}, \cite{Liao},\cite{Yee}.
Those arguments have assumed that the state of particles
is a representation of the Poincar\'{e} group
while a canonical noncommutative field theory
has symmetry of $SO(1,1)\times SO(2)$
which is a subgroup of the Poincar\'{e} group.
Recently, Chaichian et al.\cite{chaichian},
and Wess\cite{Wess00} have proposed a framework
 of making quantum theory in noncommutative spacetime
invariant under the twisted
Poincar\'{e}-Hopf algebra $\cU _{\cF}(\cP)$
using the proper twisting element
$\cF \in \cU(\cP)\otimes \cU (\cP)$, where $\cP$ is
Poincar\'{e} algebra and $\cU (\cP)$ is
its universal enveloping algebra.
In light of their work, field theory in noncommutative
spacetime can be regarded as field theory in commutative spacetime
with twisted coproduct of Poincar\'{e}-generators.
This justifies the use of the representation
of Poincar\'{e} group in studying the
noncommutative quantum field theory.
There have been some attempts
to apply this idea to the field theory with
$\Theta-$Poincar\'{e} symmetry \cite{Gonera},
conformal symmetry\cite{Matlock},\cite{Lizzi},
super conformal symmetry \cite{Choonkyu},
Galilean symmetry\cite{Sunandan},
Galileo Schr$\ddot{o}$dinger
symmetry \cite{Banerjee},
translational symmetry of $R^d$
\cite{Oeckl},
gauge symmetry\cite{Vassil},\cite{Wess0}
and diffeomorphic symmetry
\cite{Archil},\cite{Wess},\cite{Wess1}.
 To construct a consistent quantization
formalism of field theory
we need also to twist the algebra of
quantum field operators consistently.
There have been some studies
on this problem
\cite{chaichian2},\cite{Bala},\cite{Zahn}.

In this paper, we derive a twisted algebra of
creation and annihilation operators
$\{a,\ad\}$ as a basis to construct
the twisted Lorentz invariant quantum field theory.
We propose a framework to construct
a consistent quantum field theory
with this deformed algebra.
Though we focus on the quantization
of scalar field theory
in this article, we expect that
the main ideas of this article can be applied
to other field theories.
In section \ref{TwHopf},
we introduce the way to twist Hopf algebras
and target algebras,
and briefly review Chaichian et al.'s work.
In section \ref{taqo},
we introduce twisted algebra of
quantum operators,
and  we propose a framework to construct
twisted quantum field theory.
We then apply the formulation
to construct the Fock space and S-matrix
of the scalar field theory
in section ~\ref{fock}.
Finally we discuss some physical implications
of the formulation
in section \ref{discussions}.

\section{Twisted Hopf algebra of Poincar\'{e} generators}
\label{TwHopf}

Let $ \cU (\cP)$ be a universal enveloping algebra
of Poincar\'{e} Lie algebra $\cP$
and $Y (=P_\rho , M_{\mn})$
be its elements\cite{chaichian}.
They satisfy Hopf algebra properties:
\beqa
\Delta Y = Y\otimes 1 + 1 \otimes Y ,\nn\\
\epsilon (Y) = 0,\quad S(Y) = -Y,
\eeqa
where $\Delta$ is coproduct,
$\epsilon$ counit, and
$S$ antipode.
The action of $Y$
on the algebra of function space $ \cal A$
satisfies the relation
(hereafter we use Sweedler's notation
$\Delta Y =\sum Y_{(1)}\otimes Y_{(2)} $)
\beq
Y\rhd
(\phi \cdot \psi) = \sum (Y_{(1)}\rhd \phi) \cdot (Y_{(2)}
        \rhd \psi),
\label{action}
\eeq
where $\phi, \psi \in \cA$, the symbol $\cdot$ is
a multiplication in the algebra $\cA$,
and the symbol $\rhd$ denotes the action of the
Poincar\'{e} generators on the algebra $\cA$
of the complex function space.

The representation of the action of Poincar\'{e} generators
 ($P_\rho , M_{\mn}$)
on the function space is given by
\beqa
P_\rho \rhd \phi(x) &=&-i\partial_\rho \phi(x),\nn\\
M_{\mu\nu}\rhd \phi(x) &=&
-i(x_{\mu}\partial_\nu-x_{\nu}\partial_\mu) \phi(x).
\eeqa
If we have
some `twisting element' $\cF \in \cUP\otimes\cUP$,
we can generate another Hopf algebra $\cUFP$
by twisting $\cUP$ with $\cF$.
This twisting element $\cF$ must satisfy
the so called ``2-cocycle and co-unital condition"
\cite{Majid}:
\beqa
&&(\cF\otimes
1)\cdot (\Delta\otimes \text{id})\cF
=(1\otimes \cF)\cdot
(\text{id}\otimes \Delta )\cF,\nn\\
&&
(\epsilon\otimes \text{id})\cF
=1=
(\text{id}\otimes \epsilon )\cF.
\eeqa
Since the Poincar\'{e} algebra $\cal P$
has a commutative subalgbra $\{P_\rho\}$,
it is easy to construct a
twisting element $\cF$ from $P_\rho$'s :
\beqa
\cF = \exp\left(\halfi
\theta^{\alpha\beta} P_\alpha\otimes P_\beta\right).
\label{twistelement}
\eeqa
The new Hopf algebra generated
from this twisting element
is the same as the algebra part
of the original Hopf algebra
while it has different co-algebra structure.
This means that the Lie algebra commutation relations
have the same form
and the representation of Poincar\'{e} generators
remains unchanged.

The new co-product $\Delta_\cF$ has the form,
\beqa
\Delta_\cF Y = \cF \cdot \Delta Y \cdot \cF^{-1},
\label{tcoproduct}
\eeqa
with the same co-unit and antipode,
$\epsilon_\cF = \epsilon , S_\cF = S$.
Under the change of co-product,
the action  of $Y$(Eq.~(\ref{action}))
does not transform covariantly in general.
For the form of Eq.~(\ref{action}) to change covariantly,
one has to twist the target algebra $\cA$ properly.
This consistent multiplication, $*$,
of twisted algebra $\cA_\cF$ has the form
\beq
\phi * \psi = \cdot ~[\cF^{-1}\rhd (\phi\otimes\psi)].
\eeq
When $\phi, \psi \in \cA_\cF$
are the functions  of the
same spacetime coordinate $x^\mu$,
the product $*$ becomes
the well known Moyal product.
Since $P_\alpha \rightarrow -i\del_\alpha$
in this representation,
the commutation relation between
spacetime coordinates is deduced
from this $*$-product:
\beqa
 x^\mu * x^\nu &=&
 \cdot ~[e^{+\halfi\theta^{\alpha\beta}
        \del_\alpha\otimes\del_\beta}
 \rhd (x^\mu\otimes x^\nu)]\nn\\
 &=& x^\mu\cdot x^\nu + \halfi \theta^\mn,
\eeqa
which leads to the commutation relation
\beqa
 [x^\mu,x^\nu]_* &=& i \theta^\mn.
\eeqa

The above arguments imply
 that with a twisting element
 satisfying 2-cocycle condition,
one can construct a new algebra pair
$\{\cUFP , \cA_\cF\}$
from the original pair of algebras $\{\cUP,\cA\}$.
Thus, one can think of
a field theory belonging to
a class $\{\cUP,\cA\}$
in noncommutative spacetime
as a field theory belonging to
a class $\{\cUFP , \cA_\cF\}$
in commutative spacetime.
The field theory that
belongs to a class $\{\cUFP , \cA_\cF\}$
has many advantages.
An important feature of this class of theories is that
the theory has the twist-deformed Poincar\'{e} symmetry.
Moreover, the operators such as $P^2$ and $W^2$
($W_\alpha =-\half
\epsilon_{\alpha\beta\gamma\delta}
M^{\beta\gamma}P^{\delta}$)
remain Casimir operators in the twisted algebra.
So, in this framework the symmetry of the theory
is more transparent,
and one can utilize the irreducible representation
of the Poincar\'{e} algebra
in studying the noncommutative field theory.
This framework justifies the earlier studies
in noncommuatative quantum field theory
where the representation of the
Poincar\'{e} algebra has been utilized\cite{chaichian}.

\section{Twisted algebra of
creation and annihilation operators}
\label{taqo}
In the previous section,
we summarized how the conventional field theory
can be deformed to a field theory which has
a twist-deformed Poincar\'{e} symmetry.
Since the Poincar\'{e} generators act on physical
Hilbert space also, it is natural to
deform the algebra of operators
on Hilbert space covariantly
when we twist the Poincar\'{e} symmetry.
Chaichian et al. \cite{chaichian2}
and Balachandran et al.\cite{Bala}
have studied this aspect of
the noncommutative field theory
and most recently Zahn has also
investigated this aspect
(he especially considered twisting
the commutation relations of
quantum operators) \cite{Zahn}.
In this section,
we construct a noncommutative field theory
by twisting the algebra of
quantum field operators
in such a way to preserve
the action of Poincar\'{e} group on the Fock space.
In conventional field theory,
if we create/annihilate $n$-particles of
momenta $p_1,\ldots,p_n$
in a Lorentz frame,
then it implies that we create/annihilate $n$-particles
of momenta $\Lambda p_1,\ldots,\Lambda p_n$
in a Lorentz transformed frame.
Since we want to preserve this relation
in the twisted theory,
the focus in this paper
will mainly be on the relation between
the action of Poincar\'{e} group and
creation/annihilation operators.

%\subsection{}
\subsection{The Action of $\cUP$
on Algebra of Operators $\Omega$}

Let $\Omega$ denote a vector space of
selected operators
whose domain and range are
the physical Hilbert space $\cT$.
By defining the composite map of two operators
 as a multiplication
(we denote it by the symbol $\circ$),
$\Omega$ becomes an algebra
if it is closed under this multiplcation $\circ$.
In other words, for arbitrary $\Psi \in \cT,
\text{ and for all}~ a,b\in \Omega $,
the map %\circ%
\beqa
a\circ b : ~\cT &\rightarrow& \cT\nn\\
 (a\circ b)\Psi&=& a(b(\Psi)),
\eeqa
defines a multiplication
 in  $\Omega$.

We denote the action of $Y \in \cUP$
on a selected target algebra $\Omega$ as $\btr$
to distinguish it from the action $\rhd$
defined in the last section,
and let $U(\Lambda,\epsilon)$
be a Poincar\'{e} transformation in the physical
Hilbert space $\cT$
(\emph{$\Lambda$ denotes a Lorentz transformation,
$ x \rightarrow x'=\Lambda x$,
and $\epsilon$ denotes a translation,
$x\rightarrow x'=x+\epsilon$}).
From the relation,
\beqa
U(\Lambda,\epsilon)(a\Psi)&=&
U(\Lambda,\epsilon)\cdot a \cdot U^{-1}(\Lambda,\epsilon)
\left( U(\Lambda,\epsilon)\Psi \right) \nn\\
&=& a_{(\Lambda,\epsilon)}\Psi_{(\Lambda,\epsilon)},
\eeqa
where $\Psi\in \cT$ and $ a\in\Omega$,
and from the definition of $\btr$,
$a_{(\Lambda,\epsilon)} = U(\Lambda,\epsilon)\btr a$,
we have $\delta_{(\Lambda,\epsilon)}a
=-i\epsilon^c (Y_c\btr a)$
where $\epsilon^c$ are infinitesimal parameters of
Poincar\'{e} transformation and
$c$ denotes $\{{[\mn],\rho|\mu,\nu,\rho = 0,1,2,3}\}$.
These relations give the form of the action $\btr$:
\begin{subequations}
\label{blackaction}
\beqa
(Y\btr a)\Psi &=& Y\rhd(a\Psi)- a(Y\rhd\Psi),
\label{blackaction1}\\
Y\btr a &\equiv& [Y,a],
\label{blackaction2}
\eeqa
\end{subequations}
where the commutator in Eq.~(\ref{blackaction2})
is understood as in Eq.~(\ref{blackaction1}).
This operation satisfies
the properties needed to be an action:
\begin{subequations}
\label{baction}
\beqa
(Y\cdot Z) \btr a &=& Y\btr(Z\btr a),
~~~~\openone\btr a = a,~~Y\btr\openone = 0,
\label{baction1}\\
Y\btr(a\circ b)&=&\circ~[\Delta Y\btr(a\otimes b)]\nn\\
&=& \sum (Y_{(1)}\btr a)\circ(Y_{(2)}\btr b).
\label{baction2}
\eeqa
\end{subequations}
Thus, we have to twist the quantum operators properly
so as to preserve the relation Eq.~(\ref{baction}).

\subsection{Twisted Algebra of
Creation and Annihilation Operators}
As in Sec.~\ref{TwHopf}, the algebra of
quantum operators $\Omega$ has to be deformed
properly to make the form of Eq.~(\ref{baction2})
covariant.
Let this consistently twisted product of $\circ$
be denoted as $\star$.
In order to distinguish
this product from the Moyal product,
we denote
the twisted product of $\Omega_\cF$ by $\star$ and
the Moyal product of $\cA_\cF$ by $*$
throughout this paper.
The consistent form of $\star$-product is,
\beq
(a\star b)\Psi
= \circ~[\cF^{-1}\btr (a\otimes b)]\Psi,
\eeq
where $\cF$ is the same twisting element of
Eq.~(\ref{twistelement}).
The explicit form of the $\star$-product is expressed as,
\beqa
(a\star b)\Psi &=&
\sum (\cF^{-1}_{(1)}\btr a)\circ
\left((\cF^{-1}_{(2)}\btr b)\Psi\right)\nn\\
&=& \sum_{k=0}^{\infty}\fr{1}{k!} \left(-\halfi\right)^k
 \thetak\cdot (P_{\alpha 1}\btr( P_{\alpha_2}\btr
\cdots (P_{\alpha k}\btr a)\cdots)\circ
\{~(P_{\beta 1}\btr( P_{\beta_2}\btr
\cdots (P_{\beta k}\btr b)\cdots)\Psi\}\nn\\
&=&
\sum_{k=0}^{\infty}\fr{1}{k!}
\left(-\halfi\right)^k \thetak\cdot\ppa a],\cdots]
\circ\{~ \ppb b],\cdots]\Psi\}.
\label{starproduct}
\eeqa

In conventional field theory,
the scalar field operator is expressed as
(see Section 14.2 of \cite{Wald})
\beq
\hat{\phi}(x)=\int_p~
\left[~ \sigma_p(x)\cdot a(\bar{\sigma}_p) +
\bar{\sigma}_p(x)\cdot \ad(\sigma_p) \right],
~~~~~~
\int_p\equiv
\int \fr{d^3p}{(2\pi)^3 2\omega_p},
\label{scalarfield}
\eeq
where $\sigma_p$ denotes the positive frequency
 solution of
the Klein-Gordon equation, and
$a(\bar{\sigma})$ and $\ad(\sigma)$ are the
annihilation and the creation operators, respectively.
In the above relation we regard the creation and
the annihilation operators as basis operators which
act on the physical Hilbert space
and
$\sigma_p(x)$ and $\bar{\sigma}_p(x)$ as
 coefficients.
Thus the selected target algebra to be twisted
is the algebra generated by
$\{\openone, a(\sigma_p),\ad(\sigma_p),\forall p\}$.
We abbreviate $a(\sigma_p)$ and $\ad(\sigma_p)$
as $a_p$ and $\ad_p$, respectively hereafter.

The action of $P_\alpha$ on this basis operators
is represented as
\beqa
~P_\alpha\btr a_q=-q_\alpha a_q ,~&&~
~P_\alpha\btr\ad_q =+q_\alpha\cdot\ad_q,
\eeqa
which will be denoted in a simplified notation as,
\beqa
P_\alpha ~\Rightarrow ~\wt{q}_\alpha
 = \left\{
      \begin{array}{ll}
        +q_\alpha, & \text{for  } ~\ad_q ~; \\
        -q_\alpha, & \text{for  } ~a_q.
      \end{array}
    \right.
\label{tildeq}
\eeqa
From Eq.~(\ref{starproduct}), $\star$-product
between two of the creation and/or the annihilation operators
is expressed, in terms of
the conventional multiplication, as
\beqa
a_p\star~ a_q &=& e^{-\halfi p\wedge q}~a_p\cdot a_q~,
~~~~
\ad_p\star~\ad_q = e^{-\halfi p\wedge q}~\ad_p\cdot\ad_q,
\nn\\
a_p\star~\ad_q &=& e^{+\halfi p\wedge q}~a_p\cdot \ad_q~,
~~~~
\ad_p\star~a_q = e^{+\halfi p\wedge q}~\ad_p\cdot a_q.
\eeqa
This relation can be written in a compact form as
\beqa
c_p\star~c_q =
e^{-\halfi \wt p\wedge\wt q}
c_p\cdot~c_q,
~~~~~~~~~~c_p = a_p ~~\text{and/or}~~ \ad_p.
\eeqa
In Appendix~\ref{appen1}
we explicitly compute the product of
$n$ operators, $a_p ~\text{and/or}~\ad_q$, to be
\beqa
c_{q_1}\star\cdots\star ~c_{q_n}
= \cE(\wt{q}_1,\cdots,\wt{q}_n)
 ~c_{q_1}\cdots ~ c_{q_n}, ~~~~~~~
\cE (\wt{q}_1,\cdots,\wt{q}_n)=
\exp{\biggl(-\halfi \sum_{i<j}^{n}
\wt{q}_i\wedge \wt{q}_j\biggr)},
\label{nproduct}
\eeqa
where $c_q$ represents $a_q$ or $\ad_q$ and $\wt{q}$
 is defined in Eq.~(\ref{tildeq}).
%\subsubsection{properties of the twisted algebra}
Thus the $\star$-products of $n$-number of
creation or annihilation operators are just
the conventional products of the corresponding operators
multiplied by the phase factor
$\cE(\wt{q}_1,\cdots,\wt{q}_n)$.
It is worth to note that
this twisted algebra is associative as shown in
Eq.~(\ref{cnstar}),
and the complex conjugation is also compatible
with this algebra:
\beqa
(c_{q_1}\star\cdots\star~c_{q_n})^\dagger
= c_{q_n}^\dagger\star\cdots\star~c_{q_1}^\dagger,
\label{conjugation}
\eeqa
since $\bar{\cE}(\wt{q}_1,\cdots,\wt{q}_n)
=\cE(-\wt{q}_n,\cdots,-\wt{q}_1)$.

In this construction, we note that
the coefficient functions are not $\star$-producted:
\beqa
\biggl(\sum a_i(x)c_i\biggr)
\star
\biggl(\sum b_j(x)c_j\biggr)
\left\{
      \begin{array}{l}
      =\sum (a_i(x)\cdot b_j(x)) (c_i\star c_j),
         \\
       \neq \sum (a_i(x)* b_j(x)) (c_i\star c_j).
      \end{array}
    \right.
\label{dcount}
\eeqa
Since $\star$-product is deduced by
requiring covariance of the action
of Poincar\'{e} algebra
on $\Omega$, it is enough to check
this property of the $\star$-product
by evaluating the action
on the free scalar field case.
Specifically,
let $U_\Lambda$ be a Lorentz transformation
which acts only on the operator part,
but not on the spacetime part, then
\beqa
\hat{\phi}'(x)
&=&
U_\Lambda\cdot\hat{\phi}(x)\cdot U^{-1}_\Lambda\nn\\
&=&\int_p
\left[~ e^{ip\cdot x}\cdot U_\Lambda\cdot
a_p\cdot U^{-1}_\Lambda
+e^{-ip\cdot x}\cdot U_\Lambda\cdot
\ad_p\cdot U^{-1}_\Lambda~\right]\nn\\
&=&\int_p
\left[~ e^{ip\cdot x}\cdot a_{\Lambda p}
+e^{-ip\cdot x}\cdot \ad_{\Lambda p}~\right]\nn\\
&\equiv&\int_p
\left[~ e^{ip\cdot \Lambda x}\cdot a_p
+e^{-ip\cdot \Lambda x}\cdot \ad_p~\right]\nn\\
&=&
\hat{\phi}(\Lambda x),
\eeqa
where the use has been made of the fact that
$p\cdot x = \Lambda p\cdot \Lambda x$.
This leads to
\beqa
Y\btr\hat{\phi}(x)&\approx& Y\rhd\hat{\phi}(x).
\eeqa
This suggests other possibilities of twisting
the algebra of quantum field operators.
Since quantum field operators are functions of
spacetime and are also operators in the Hilbert space,
one may twist the quantum operator part
as indicated in \cite{Fiore0},
and carried out in this paper,
or twist the spacetime function part
of the field operators as has been done by
Chaichian et al.\cite{chaichian2},
and Zahn\cite{Zahn}.
Or one may twist both the quantum operator part
and the spacetime function part as has been
done by Balachandran et al.\cite{Bala},
which will result different S-matrix
due to the cancellation of the effects of
the two twist operations.
We have chosen in this paper to twist
the algebra of quantum operator part of
the field operators,
which gives the same S-matrix as those of
earlier studies\cite{Filk}, as will be shown
in the next section.

Incidentally, our $\star$-product has the
same expression as that of ref.~\cite{chaichian2}
 when it is performed on the
scalar field operators:
\beqa
&&\hat{\phi}(x)\star \hat{\phi}(y)\nn\\
&=&\int_{p,q}\left[~
e^{ipx}e^{iqy}(a_p\star a_q)
+ e^{ipx}e^{-iqy}(a_p\star \ad_q)\right.
\left.+ e^{-ipx}e^{iqy}(\ad_p\star a_q)
+ e^{-ipx}e^{-iqy}(\ad_p\star \ad_q)
~\right]\nn\\
&=&\int_{p,q}\left[~
e^{-\halfi p\wedge q}e^{ipx}e^{iqy}(a_p\cdot a_q)
+e^{+\halfi p\wedge q}e^{ipx}e^{-iqy}(a_p\cdot \ad_q)
\right.\nn\\
&&~~~~~~~~~~~~~~~~~~~~~~~~~~~~~~~~~~\left.
+e^{+\halfi p\wedge q}e^{-ipx}e^{iqy}(\ad_p\cdot a_q)
+e^{-\halfi p\wedge q}e^{-ipx}e^{-iqy}(\ad_p\cdot \ad_q)
\right]\nn\\
&=&e^{\halfi\del_x\wedge\del_y}
\left(\hat{\phi}(x)\cdot\hat{\phi}(y)\right)\nn\\
&\equiv&
\hat{\phi}(x)\star \hat{\phi}(y)|_{\text{Chaichian et al.}.}
\eeqa
This shows that our expression for the $\star$-product
corresponds to the momentum space representation of
Chaichian et al.'s $\star$-product for the free
scalar field case.

\section{Physical Fock space and S-matrix}
\label{fock}
In the previous section,
we have constructed twisted algebra
of quantum operators.
The physical quantities such as S-matrix
must also be written in twist covariant form.
Since the physical quantities
 can be expressed as a sum of products of
creation and annihilation operators
in the conventional field theory,
the physical quantities in the twisted theory
must be expressed as the same quantities
with the conventional product replaced by
the $\star$-product.
Thus we can consistently
construct noncommutative field theory
by twisting the conventional field theory.

\subsection{Twisted Fock space}

\subsubsection{Commutative case}
\label{cqft}
Let $\cH$ denote a one particle Hilbert space of
scalar field theory.
Then the Fock space of this theory
 can be written as
$T(\cH) = C\oplus[\bigoplus_{n=0}^{\infty}\cH_S^n]$,
 $\cH_S^n\equiv\bigotimes_{S}^n\cH$,
 where subscript 'S' denotes symmetrization
 and $\bigotimes^n$ denotes n'th order
 tensor product $\cH\otimes\cdots\otimes\cH$.
The action of creation and annihilation operators
on the normalized $n$-particle state $\phiqn\in \cH_S^n$,
is expressed as
%(we use the notation of ref.~\cite{Weinberg}),
\beqa
\label{annihil1}
\ad_q\phiqn &=&\ket{q,q_1,\cdots,q_n},\nn\\
a_q\phiqn &=& \sum_{k=1}^{n}\delta(q-q_k)
\ket{q_1,\cdots, q_{k-1},q_{k+1},\cdots q_n}.
\eeqa
They satisfy the fundamental commutation relations,
\beqa
[\ad_p,\ad_q]=0,&&[a_p,a_q] = 0,\nn \\
~ [ a_p,\ad_q] &=&\delta (p-q).
\label{comrelation}
\eeqa
Any elements $\phiqn$ in the Fock space can be
obtained by successive operations of creation
operators to the vacuum state $\ket{0}$
($ a_q\ket{0} = 0 $, for all momentum $q$),
\beq
\phiqn = \ad_{q_1}\cdots~ \ad_{q_n}\ket{0}.
\label{phiqn}
\eeq

\subsubsection{Noncommutative case}

In noncommutative spacetime,
$\phiqn$ in Eq.~(\ref{phiqn}) is mapped into
\beq
\psiqn = \ad_{q_1}\star~\ad_{q_2}\star
\cdots\star~\ad_{q_n}\ket{0},
\eeq
which will be called twisted $n$-state
rather than $n$-particle state.
If the vacuum state is defined as
$\ket{0}_\star=\ket{0}$,
the twisted 1-state is the same as
the one particle state
in the conventional quantum field theory:
\beqa
\ket{q}_\star=\ad_q\star\ket{0}_\star =
(\ad_q\star\openone)\ket{0}=\ad_q\ket{0} = \ket{q}.
\eeqa

This definition of twisted $n$-state seems very natural,
but due to the noncommutativity of the $\star$-product,
the state vector $\psiqn$ is not symmetric
under the permutation
of $(q_1,q_2,\cdots,q_n)$.
The explicit form of $\psiqn$ is
\beqa
\psiqn &=& \cE (q_1,\cdots,q_n)\phiqn,
~~~~~
\cE (q_1,\cdots,q_n)=
\exp{\biggl(-\halfi \sum_{i<j}^{n} q_i\wedge q_j\biggr)}.
\eeqa
The state $\phiqn$ is symmetric
under any permutation of $(q_1,\cdots,q_n)$,
but the phase factor $\cE (q_1,\cdots,q_n)$
is not symmetric in general for $n\geqq 2$.
%(except for case of which all momentums are equal).
Since the phase factor has unit norm ($|\cE|=1$),
$\psiqn$ and $\phiqn$ are in the same ray
of the physical Hilbert space.

Some properties of the phase factor $\cE$
are listed in Appendix \ref{appen2}.
In this new algebra the creation and the annihilation
operators do not satisfy the fundamental commutation
relation Eq.~(\ref{comrelation}),
rather, they satisfy
(the same form of relations appear in \cite{Bala}),
\beqa
\ad_p\star~\ad_q &=&
e^{-i p\wedge q}\cdot\ad_q\star~\ad_p,
~~~~~~
a_p\star ~a_q =
e^{-i p\wedge q}\cdot a_q\star ~a_p,\nn\\
a_p\star ~\ad_q &=&
 e^{+i p\wedge q}\cdot \ad_q\star ~a_p
 +\delta (p-q).
\label{noncomrelation}
\eeqa
The action of creation and annihilation operators
on the state $\psiqn$ gives
% in different form,
\beqa
\label{annihil2}
\ad_q\star\psiqn
&=&\ket{q,q_1,\cdots, q_n}_\star\nn\\
a_q\star\psiqn&=&
\sum_{k=1}^{n}
\delta(q-q_k)e^{iq\wedge(q_1+\cdots+q_{k-1})}
\ket{q_1,\ldots,q_{k-1},q_{k+1},\ldots,q_n}_\star.
\eeqa

%As we observe in
%Eq.~(\ref{noncomrelation}),~(\ref{annihil2}),
%we have to be careful not to change algebra
%of derived results.
%If we change the algebra of Eq.~(\ref{comrelation})
%to $\star$-product,
%we would get $[a_p,a_q]_\star = 0$, etc.,
%%In this viewpoint, we can not say that
%the commutation relation
%in Eq.~(\ref{comrelation}) is fundamental.
From these twisted states $\psiqn$,
we define twisted Fock space as
$T_{\cF}(\cH) =   C\oplus[\bigoplus_{n=0}^{\infty}\cH_{\cF}^n]$,
( $\cH_\cF^n\equiv\bigotimes^n\cH$ ).
\\
Using the above twisting process,
it is natural to define the total number operator as
\beq
N_\cF= \sum_{k}\ad_k\star a_k.
\eeq
This number operator satisfies
\beq
[N_\cF,\ad_q]_\star=\ad_q ~~,~~  [N_\cF,a_q]_\star=- a_q.
\eeq
The state $\psiqn$
has the same eigenvalue
for the number operator
as that of the state $\phiqn$
in the conventional theory, i.e,
the eigenvalue equation,
\beqa
N\phiqn &=& n(q_1,\cdots,q_n)\cdot \phiqn,\nn
\eeqa
leads to
\beqa
N_\cF\star \psiqn
 &=& \sum_k\ad_k\star a_k\star
 \ad_{q_1}\star\cdots\star\ad_{q_n}\ket{0}\nn\\
&=&
\sum_{k}\cE(k,-k,q_1,\cdots,q_n)\cdot N_k\cdot\phiqn
\nn\\
&=&\cE(q_1,\cdots,q_n)\cdot ( N \phiqn)\nn\\
&=& n(q_1,\cdots,q_n)\cdot\psiqn,
\label{Npsi}
\eeqa
where we have used the relation Eq.~(\ref{facphase}).

The Hamiltonian for the free scalar field theory
has the form,
\beqa
H_\cF = \sum_k \omega_k \cdot\ad_k \star a_k,
~~\text{where}~~\omega_k =\sqrt{k^2+m^2}.
\eeqa
Thus as in the number operator case,
the state $\psiqn$
has the same energy eigenvalues
as the state $\phiqn$
for the free Hamiltonian
of the commutative scalar field theory.
%We consider self interacting Hamiltonian
%when we twist S-matrix.

\subsection{S-matrix}
\label{secsmatrix}
%\subsubsection{Construction of the S-matrix}
%Since our proposal for the S-matrix is
%changing the conventional operator product to
%$\star$-product,
%the twisted S-matrix can be written as,
By properly twisting the algebra of the quantum
operators we have the expression
for the S-matrix in the noncommutative field theory:
\beqa
S_{\star} &=&
T\exp{\left(-i\int d^4x~ \cH_I^\star(x)\right)}\\
&=&
\sum_{k=0}^{\infty}\fr{(-i)^k}{k!}
\int d^4x_1\cdots d^4x_k~
T\left\{\cH_I^\star(x_1)\star\cdots\star
\cH_I^\star(x_k)\right\},
\nn
\eeqa
where
$\cH_I^\star$ is the interacting part
of the Hamiltonian and $T$ denotes the time ordering.
Since we can not define interaction Hamiltonian $\cH_I$
in space-time noncommutative case in general,
we assume the space-space noncommutative case only
in this section.
For the $g\phi^n(x)$ theory,
the element of S-matrix can be calculated by using
the properties of the phase factor $\cE$,
given in Appendix
~\ref{appen2}. If we use the abbreviation
\beqa
\phi_{in}(x)=
\int_q\left[\sigma_{-q}(x)a_q
+\sigma_q(x)\ad_q\right]
\equiv\sum_{c_q=a_q,\ad_q}
\int_q~ c_q\cdot\sigma_{\wt{q}}(x),
\eeqa
the interaction Hamiltonian can be written as
\beqa
\cH_I^\star(x)&=&
g\int_{q_1}\cdots\int_{q_n}
(c_{q_1}\star\cdots\star~c_{q_n})\cdot
\sigma_{\wt{q}_1}(x)\cdots\sigma_{\wt{q}_n}(x)\nn\\
&\equiv&
g\sum_{c_Q}\int_Q c_Q^\star\cdot\sigma_{\wt{Q}}(x),
\eeqa
where
$\displaystyle
\int_Q\equiv\int_{q_1}\cdots\int_{q_n},~
c_Q^\star\equiv c_{q_1}\star\cdots\star~c_{q_n},~
\sigma_{\wt{Q}}(x)
\equiv\sigma_{\wt{q}_1}(x)\cdots\sigma_{\wt{q}_n}(x)
,~\sum_{c_Q}\equiv
\sum_{c_{q_1}=a_{q_q},\ad_{q_1}}
\cdots\sum_{c_{q_n}=a_{q_n},\ad_{q_n}}.$

The twisted S-matrix is then expressed as
\beqa
S_\star
&=&\sum_{k=0}^{\infty}\fr{(-i)^k}{k!}
\int d^4x_1\cdots d^4x_k~
T\left\{\cH_I^\star(x_1)\star\cdots\star
\cH_I^\star(x_k)\right\}\nn\\
&=&\sum_{k=0}^{\infty}(-i)^k
\int d^4x_1\cdots d^4x_k~
\theta(x_1,\cdots,x_k)
\cH_I^\star(x_1)\star\cdots\star
\cH_I^\star(x_k)\nn\\
&=&\sum_{k=0}^{\infty}(-ig)^k
%\sum_{c_{Q_1}\cdots ~c_{Q_k}}
\int_{Q_1}\cdots\int_{Q_k}
~\sum_{c_{Q_1}\cdots ~c_{Q_k}}
c_{Q_1}^\star\star\cdots \star~c_{Q_k}^\star
\biggl(
\int d^4x_1\cdots d^4x_k~
\theta(x_1,\cdots,x_k)
\sigma_{\wt{Q}_1}(x_1)\cdots\sigma_{\wt{Q}_k}(x_k)
\biggr)\nn\\
&=&\sum_{k=0}^{\infty}(-ig)^k
\int_{Q_1}\cdots\int_{Q_k}
\sum_{c_{Q_1}\cdots ~c_{Q_k}}
\cE(\wt{Q}_1,\cdots,\wt{Q}_k)
~c_{Q_1}\cdots ~c_{Q_k}\cdot
\wt{\Theta}(\wt{Q}_1,\cdots,\wt{Q}_k),
\label{smatrix}
\eeqa
where
\beqa
\wt{\Theta}(\wt{Q}_1,\cdots,\wt{Q}_k)
=\int d^4x_1\cdots d^4x_k~
\theta(x_1,\cdots,x_k)
\sigma_{\wt{Q}_1}(x_1)\cdots\sigma_{\wt{Q}_k}(x_k).
\eeqa

In the limit $\theta\rightarrow 0$,
this S-matrix reduces to the one
in the commutative case:
\beqa
S_\star\rightarrow
S&=&\sum_{k=0}^{\infty}(-ig)^k
\int_{Q_1}\cdots\int_{Q_k}
~\sum_{c_{Q_1}\cdots ~c_{Q_k}}
c_{Q_1}\cdots ~c_{Q_k}\cdot
\wt{\Theta}(\wt{Q}_1,\cdots,\wt{Q}_k)\nn\\
&=&\sum_{k=0}^{\infty}(-ig)^k \int_{Q_1}\cdots\int_{Q_k}
\sum_{c_{Q_1}\cdots ~c_{Q_k}} S^k(\wt{Q}_1,\cdots,\wt{Q}_k),
\label{limitS} \eeqa where $S^k(\wt{Q}_1,\cdots,\wt{Q}_k)
=c_{Q_1}\cdots ~c_{Q_k}\cdot
\wt{\Theta}(\wt{Q}_1,\cdots,\wt{Q}_k)$
corresponds to the momentum space representation
of $k$-th order term of the S-matrix
in the conventional field theory.
Eq.~(\ref{smatrix}) and Eq.~(\ref{limitS})
show the relation between the $S_\star$-matrix
and the S-matrix of the
corresponding commutative theory.
For the S-matrix element
$_\star\bra{\beta}S_\star\ket{\alpha}_\star$, where
$\ket{\alpha}_\star(\ket{\beta}_\star)$
denotes `$in$' twisted
$n(m)$-state, we have
\begin{subequations}
\label{2smatrix}
\beqa
\label{smatrixba}
_\star\bra{\beta}S_\star\ket{\alpha}_\star
&=&\sum_{k=0}^{\infty}(-ig)^k
\int_{Q_1}\cdots\int_{Q_k}
\sum_{c_{Q_1}\cdots ~c_{Q_k}}
\cE(\wt{\beta},\wt{Q}_1,\cdots,\wt{Q}_k,\wt{\alpha})
~\bra{\beta}c_{Q_1}\cdots ~c_{Q_k}
\ket{\alpha}\cdot
\wt{\Theta}(\wt{Q}_1,\cdots,\wt{Q}_k)\nn\\
&=&
\cE(\wt{\beta},\wt{\alpha})
\sum_{k=0}^{\infty}(-ig)^k
\int_{Q_1}\cdots\int_{Q_k}
\sum_{c_{Q_1}\cdots ~c_{Q_k}}
\cE(\wt{Q}_1,\cdots,\wt{Q}_k)
~\bra{\beta}S^k(\wt{Q}_1,\cdots,\wt{Q}_k)
\ket{\alpha}
~~~~(\because\wt{\alpha}+\wt{\beta}=0)
\nn\\
&=&
\cE(-\beta,\alpha)\bra{\beta}S_\star\ket{\alpha},
\eeqa
where the momenta $\wt{\alpha},\wt{\beta}$
are related to those of
$\alpha =\ket{\alpha_1,\alpha_2,\cdots},
\beta =\ket{\beta_1,\beta_2,\cdots}$
as shown in Fig.~\ref{feynman}, and
\begin{figure}
%\begin{tabular}{cc}
\includegraphics[width=0.7\linewidth]{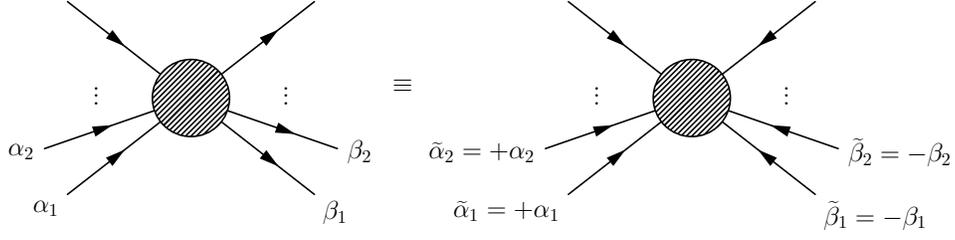}
%\end{tabular}
\caption{Illustration of the notation
for momenta $\wt{\alpha}$ and $\wt{\beta}$}
\label{feynman}
\end{figure}
\beqa
\label{smatrixfilk}
\bra{\beta}S_\star\ket{\alpha}&=&
\sum_{k=0}^{\infty}(-ig)^k
\int_{Q_1}\cdots\int_{Q_k}
\sum_{c_{Q_1}\cdots ~c_{Q_k}}
\cE(\wt{Q}_1,\cdots,\wt{Q}_k)
~\bra{\beta}S^k(\wt{Q}_1,\cdots,\wt{Q}_k)
\ket{\alpha}.
\eeqa
\end{subequations}

It can be easily shown that
the result (\ref{smatrixfilk}) is the same as
that of Filk \cite{Filk}.
Since the quantities
$\bra{\beta}c_{Q_1}\cdots ~c_{Q_k}
\ket{\alpha}$ contain the energy-momentum
conservation delta functions
$\delta(\wt{Q}_1)\cdots\delta(\wt{Q}_k)$,
$\cE(\wt{Q}_1,\cdots,\wt{Q}_k)$
in Eq.~(\ref{smatrixfilk})
can be written as
$\cE(\wt{Q}_1)\cdots\cE(\wt{Q}_k)$, and
each of these $\cE(\wt{Q}_i)$ gives the phase factor
at each vertex in the Feynman diagram.

%\beqa
%\bra{\beta}S_\star\ket{\alpha}&=&
%\sum_{k=0}^{\infty}(-ig)^k
%\int_{Q_1}\cdots\int_{Q_k}
%\sum_{c_{Q_1}\cdots ~c_{Q_k}}
%\cE(\wt{Q}_1)\cdots\cE(\wt{Q}_k)
%~\bra{\beta}S^k(\wt{Q}_1,\cdots,\wt{Q}_k)
%\ket{\alpha}\nn\\
%&=&\bra{\beta}
%\sum_{k=0}^{\infty}(-ig)^k
%\int d^4x_1\cdots d^4x_k~
%\theta(x_1,\cdots,x_k)
%\left(\int_{Q_1}\sum_{c_{Q_1}}\cE(\wt{Q}_1)
%c_{Q_1}\sigma_{\wt{Q}_1}(x)\right)
%\cdots
%\left(\int_{Q_k}\sum_{c_{Q_k}}\cE(\wt{Q}_k)
%c_{Q_k}\sigma_{\wt{Q}_k}(x)\right)
%\ket{\alpha}\nn\\
%&=&
%\bra{\beta}
%\sum_{k=0}^{\infty}\fr{(-ig)^k}{k!}
%\int d^4x_1\cdots d^4x_k
%~T\left\{\cH_I^\star(x_1)\cdots
%\cH_I^\star(x_k)\right\}
%\ket{\alpha}.
%\eeqa

\subsection{Statistics of indistinguishable particles}
\label{statistics}
%The argument of this section is motivated
%by \cite{Weinberg},\cite{dewitt}.\\
The unit matrices in the space of $N$-particle state can be
expressed as
\beqa
&&\openone_D =\sum_{\gamma}\ket{\gamma}_D\bra{\gamma}_D
,\nn\\
&&\openone_C =
\frac{1}{N!}\sum_{\gamma}\ket{\gamma}\bra{\gamma},\nn\\
&&\openone_\star =
\frac{1}{N!}\sum_{\gamma}\ket{\gamma}_\star~_\star
\bra{\gamma}~~~(~\equiv \openone_C),
\label{unitmatrix}
\eeqa
where the subscript $C$($D$)
denotes indistinguishable(distinguishable) state
in the conventional field theory and
$\star$ denotes the noncommutative case.
%, and
%$D$ means distinguishable particle states.
We must be careful in the order of
momenta in kets and bras because
the twisted $n$-state is not symmetric under
the permutation of the momenta.
From  Eq.~(\ref{conjugation})
we find $\ket{\bf
k_1,\cdots,\bf k_N}_\star^\dagger =~
_\star\bra{\bf k_N,\cdots,\bf
k_1}$,
i.e., $\ket{\gamma}_\star\equiv \ket{\bf k_1,\cdots,\bf
k_N}_\star$ and
 $\bra{\gamma}_\star\equiv~_\star\bra{\bf k_N,\cdots,\bf
k_1}$.

Let $\alpha(\beta)$ denote
the free $N$-particle \textit{in(out)}
twisted $n(m)$-state respectively, and $\mcP$
denotes arbitrary permutation.
Then, for spatial dimensions $d\geq3$,
(See \cite{Weinberg},\cite{dewitt}), we have
\beqa
_\star\braket{\beta}{\alpha}_\star
&=&\cE(-\beta,\alpha)\braket{\beta}{\alpha}_C
~=~\cE(-\beta,\alpha)\sum_{\mcP}C_{\mcP}
\braket{\mcP(\beta)}{\alpha}_D,~~~~~~~~~~~~~~~~~~~~~~
\eeqa
where $C_{\mcP}$ are complex constants.
From the form of $\openone_\star$ and $\openone_D$ in
Eq.~(\ref{unitmatrix})
we finally obtain the relation,
\beqa _\star\braket{\beta}{\alpha}_\star
&=&\fr{1}{N!}\sum_{\gamma}{}_\star\braket{\beta} {\gamma}_\star
{}_\star\braket{\gamma}{\alpha}_\star\nn\\
&=&\fr{1}{N!}\sum_{\gamma}\cE(-\beta,\gamma)\cE(-\gamma,\alpha)
\sum_{\mcP',\mcP''}C_{\mcP'}C_{\mcP''}
\braket{\mcP'(\beta)}{\gamma}_D
\braket{\mcP''(\gamma)}{\alpha}_D\nn\\
&=&\cE(-\beta,\alpha)
\fr{1}{N!}
\sum_{\mcP',\mcP''}C_{\mcP'}C_{\mcP''}\sum_{\gamma}
\braket{\mcP'(\beta)}{\gamma}_D
\braket{\mcP''(\gamma)}{\alpha}_D\nn\\
&=&\cE(-\beta,\alpha)
\fr{1}{N!}\sum_{\mcP',\mcP''}C_{\mcP'}C_{\mcP''}\sum_{\gamma}
\braket{\mcP''\mcP'(\beta)}{\mcP''(\gamma)}_D
\braket{\mcP''(\gamma)}{\alpha}_D\nn\\
&=&\cE(-\beta,\alpha)
\fr{1}{N!}\sum_{\mcP',\mcP''}C_{\mcP'}C_{\mcP''}
\braket{\mcP''\mcP'(\beta)}{\alpha}_D.
\eeqa
Since $\cE(-\beta,\alpha)\neq 0$, we have
\beqa
\sum_{\mcP}C_{\mcP}
\braket{\mcP(\beta)}{\alpha}_D
&=&
~\fr{1}{N!}\sum_{\mcP',\mcP''}C_{\mcP'}C_{\mcP''}
\braket{\mcP''\mcP'(\beta)}{\alpha}_D\nn\\
\Rightarrow~~~~~
C_{\mcP'\mcP''}&=&C_{\mcP'}\cdot C_{\mcP''},
\eeqa
which is the one dimensional representation
of the permutation group as we have
in the conventional field theory case.
Consequently,
we have the same statistics
for indistinguishable particles
in the noncommutative field theories
as in the corresponding commutative case
\cite{Fiore0},\cite{Tureanu}.

\section{Summary and discussions}
\label{discussions}

We have constructed noncommutative
 quantum field theory
by properly twisting the algebra of
creation and annihilation operators.

As mentioned in Section~\ref{fock}
the twisted $n$-state
is not symmetric under the permutation
of its momenta.
If we permute its momenta
the state changes by a phase factor $\cE$
which has unit norm.
Thus $\phiqn$ and $\psiqn$
are in the same ray in Hilbert space,
and as we have shown in section~\ref{fock}
the phase factor is
always factorized out of the S-matrix element.
Moreover, the states $\psiqn$ and $\phiqn$ have
the same eigenvalues
for the corresponding number operators.
Hence we have the same physics whether
we use $\psiqn$ or $\phiqn$ as a basis for
\textit{in}/\textit{out} states
(for example, twisted Lorentz transformation
changes only the phase factor
of the twisted $n$-states).
Hence {\it  we can define $n$-particle state
as an equivalence class of this twisted $n$-states.}
%differ by phase factor.\\

We have shown that
S-matrix elements differ by phase factors
from the S-matrix elements of the conventional theory.
In SSNC quantum field theory
it gives the phase factor to every vertex
in the Feynman diagram.
This phase factor is the same as that in \cite{Filk},
thus justifying the results of Filk.
The expression of S-matrix in this paper
is manifestly twist Lorentz covariant,
except that time ordering may break
this symmetry since $\star$-commutator of two interacting
Hamiltonians separated by space-like distance
does not vanish in general, i.e.,
\beqa
\left[\cH_I^\star(x),\cH_I^\star(y)\right]_\star
\neq 0, ~~~\text{for } (x-y)^2 < 0.
\label{brokelocal}
\eeqa
This possible violation of
locality % i.e. Eq.~(\ref{brokelocal})
is known to be inherited
from the presence of spacetime noncommutativity.
Some authors have argued that
the micro causality is satisfied
in the SSNC case,% \cite{chaichian2},
 while it is violated in STNC case \cite{Ma}.
It appears that further studies are needed
to have a consistent STNC
field theory.

We have observed that
the statistics of indistinguishable particles
do not change in the properly twisted formulation
of the noncommutative field theory.
It reflects
the co-homological properties of
the phase factors,
i.e. $\cE(-\beta,\gamma)\cE(-\gamma,\alpha)
=\cE(-\beta,\alpha)$.

In summary,
by careful construction of quantum field theory
using the twisted algebra
applied to the quantum operator space,
we have constructed the S-matrix
of the interacting noncommutative scalar field theory,
and the result is shown to be consistent with
earlier ones \cite{Filk}.
We hope that this formulation can be generalized
to the case of more general
noncommutative field theories.

\begin{appendix}
\section{Explicit calculations
of $\star$-product
of creation and annihilation operators}
\label{appen1}
In this appendix,
we derive the explicit form of $\star$-product
of the creation and annihilation operators
in terms of the conventional products.
%\newtheorem*{lem}{Lemma}
%\begin{lem}

\textbf{1}. The $\star$-product of $(c_{p_1}\cdots~ c_{p_k})$
and $(c_{q_1}\cdots~ c_{q_l})$ is given by
\beqa
 (c_{p_1}\cdots~ c_{p_k})\star(c_{q_1}\cdots~ c_{q_l})
=e^{-\halfi \wt{P}\wedge \wt{Q}} (c_{p_1}\cdots~ c_{p_k}
c_{q_1}\cdots~ c_{q_l}),
 \eeqa
where $\wt{P}=\sum \wt{p}_i ~~~\text{and}~~
\wt{Q}=\sum \wt{q}_j$.
\begin{proof}
\beqa
P_\alpha\btr(c_{q_1}\cdots~ c_{q_l})
&=& \cdot~[\Delta^{(l)} P_\alpha\btr
(c_{q_1}\otimes\cdots\otimes c_{q_l})]\nn\\
&=&\cdot~\sum_{i=1}^l
\left[c_{q_1}\otimes\cdots
\otimes(P_\alpha\btr c_{q_i})\otimes
\cdots\otimes c_{q_l}\right]\nn\\
&=&\sum_i(\wt{q_i})_\alpha
(c_{q_1}\cdots~c_{q_i}\cdots~ c_{q_l})\nn\\
&=&\wt{Q}_\alpha
(c_{q_1}\cdots~ c_{q_l}),\nn
\eeqa
and since $(c_{p_1}\cdots~ c_{p_k})
,(c_{q_1}\cdots~ c_{q_l})\in \Omega$, we have
\beqa
(c_{p_1}\cdots~ c_{p_k})\star(c_{q_1}\cdots~ c_{q_l})
&=&\cdot \left\{
e^{-\halfi \theta^{\alpha\beta}
P_\alpha\otimes P_\beta}\btr
\left[(c_{p_1}\cdots~ c_{p_k})\otimes
(c_{q_1}\cdots~ c_{q_l})\right]
\right\}\nn\\
&=&e^{-\halfi \theta^{\alpha\beta}
\wt{P}_\alpha \wt{Q}_\beta}
(c_{p_1}\cdots~ c_{p_k}c_{q_1}\cdots~ c_{q_l}).\nn
\eeqa
\end{proof}

%\newtheorem*{thrm}{Theorem}
%\label{cnstartheorem}
%\begin{thrm}
\textbf{2}. For all natural number $n$, we have
\beqa
\label{cnstar}
c_{q_1}\star\cdots\star ~c_{q_n}
= \cE(\wt{q}_1,\cdots,\wt{q}_n)
 ~c_{q_1}\cdots ~ c_{q_n},~~~~~~~~~
\cE (\wt{q}_1,\cdots,\wt{q}_n)=
\exp{\biggl(-\halfi \sum_{i<j}^{n}
\wt{q}_i\wedge \wt{q}_j\biggr)}.
\eeqa
%\end{thrm}

\begin{proof}
(\ref{cnstar}) can be shown
by mathematical induction:\\
i)  It holds for $k=1,2$.\\
ii) Suppose that this relation
holds for $k,l\leq 1,\ldots,n ~~ (n\geq 2)$,
then for $n+1\leq N=k+l\leq 2n$,
\beqa
(c_{q_1}\star\cdots\star c_{q_k})\star
(c_{q_{k+1}}\star\cdots\star c_{q_{N}})
&=&\cE(\wt{q}_1,\cdots,\wt{q}_k)
\cE(\wt{q}_{k+1},\cdots,\wt{q}_{N})\cdot
(c_{q_1}\cdots c_{q_k})\star
(c_{q_{k+1}}\cdots c_{q_{N}})\nn\\
&=&\cE(\wt{q}_1,\cdots,\wt{q}_k)
\cE(\wt{q}_{k+1},\cdots,\wt{q}_N)\cdot
e^{-\halfi \wt{Q}_1\wedge \wt{Q}_2}
(c_{q_1}\cdots c_{q_k})\cdot
(c_{q_{k+1}}\cdots c_{q_N})\nn\\
&=&\cE(\wt{q}_1,\cdots,\wt{q}_k,
\wt{q}_{k+1},\cdots,\wt{q}_N)\cdot
(c_{q_1}\cdots c_{q_k}\cdot
c_{q_{k+1}}\cdots c_{q_N})\nn\\
&=&\cE(\wt{q}_1,\cdots,\wt{q}_N)\cdot
(c_{q_1}\cdots c_{q_N}),
\eeqa
where $Q_1=q_1+\cdots+q_k,
Q_2=q_{k+1}+\cdots+q_N$.
Hence, (\ref{cnstar}) is satisfied for all natural number
$n$, and this equation also proves the associativity of $\star$-product.
\end{proof}

Using the above theorem we find the action of
creation and annihilation operators
on the twisted states to be,
\begin{subequations}
\beqa
\ad_q\star\psiqn
&=&\cE({q,q_1,\ldots,q_n})\cdot
\ket{q,q_1,\cdots, q_n}
=\ket{q,q_1,\cdots, q_n}_\star,
\label{annhila1}
\\
a_q\star\psiqn&=&
\cE(-q,q_1,\cdots,q_n)
\cdot a_q \ad_{q_1}\cdots \ad_{q_n}\ket{0}\nn\\
&=&\cE(-q,q_1,\cdots,q_n)
\sum_{k=1}^{n}\delta(q-q_k)
\ket{q_1,\ldots,q_{k-1},q_{k+1},\ldots,q_n}\nn\\
&=&
\sum_{k=1}^{n}\delta(q-q_k)
\cE(-q,q_1,\cdots,q_{k-1},q,q_{k+1},\cdots,q_n)
\ket{q_1,\ldots,q_{k-1},q_{k+1},\ldots,q_n}\nn\\
&=&
\sum_{k=1}^{n}
\delta(q-q_k)e^{iq\wedge(q_1+\cdots+q_{k-1})}
\ket{q_1,\ldots,q_{k-1},q_{k+1},\ldots,q_n}_\star.
\label{annihila2}
\eeqa

\end{subequations}

\section{Properties of phase factor $\phase$}
\label{appen2}
In this appendix we summarize
the useful properties of the phase factor:
\beq
\phase ~=~
\exp{\biggl(-\halfi\sum_{i<j}^{n}q_i\wedge q_j
\biggr),}\nn
\eeq
where for $n=1$, we define $\cE(q)= 1$. %since $\cE(q)=\cE(q,k,-k) = 1$.
From direct calculation we find
\beqa
 \cE(k,-k)&=&1,~~
 \cE(q,k,-k)=1,
\eeqa
and since the phase factor is quadratic in $q$'s,
we have
\beqa
 \phase &=& \cE(-q_1,\cdots,-q_n).
\eeqa
The phase factor is invariant under the cyclic
permutations $\mcP$ of $(q_1,\ldots,q_n)$ if
$\sum q_k = 0$, i.e.,
\beqa
 \phase = \cE(q_{\mcP 1},\cdots,q_{\mcP n}),
~~~\text{for } \sum q_k = 0.
 \label{B3}
\eeqa
It also has the following property:
\beqa
\cE(p_1,\cdots,p_n,q_1,\cdots,q_m)
=e^{-\halfi(p_1+\cdots+p_n)
\wedge(q_1+\cdots+q_m)}
\cE(p_1,\cdots,p_n)
\cE(q_1,\cdots,q_m).
\label{B4}
\eeqa

If the sum of $m$-consecutive $q$'s is zero,
the phase factor is factorized into the product
of two factors:
\beqa
\phase &=&\cE(q_1,\ldots,q_k,q_{k+m+1},\ldots,q_n)
\cdot\cE(q_{k+1},\ldots,q_{k+m}),
~~~\text{if }\sum_{i=k+1}^{k+m}q_i=0~.\nn
\eeqa
This is a direct consequence of
(\ref{B3}) and (\ref{B4}). For $m=2$ case, we have
\beqa
%&\Longrightarrow&~~~~
\cE(q_1,\ldots,p,-p,\ldots,q_n)=\phase.
\label{facphase}
\eeqa
From the above relations, we have
\beqa
%\cE(q_1,\cdots,q_n,-q_n,\cdots,-q_1)
%&=&1=
%\cE(q_1,\cdots,q_n)\cE(-q_n,\cdots,-q_1)
%=\cE(q_1,\cdots,q_n)\cE(q_n,\cdots,q_1)\nn\\
%\Longrightarrow~~~~
\bar{\cE}(q_1,\cdots,q_n)
=\cE(q_n,\cdots,q_1)
=\cE(q_{\mcP n},\cdots,q_{\mcP 1}),
\eeqa
where $\mcP$ is a cyclic permutation.

From these results we can derive the properties of $\cE$
given in ref.~\cite{Filk}:
\begin{subequations}
\label{filkeq}

\beqa
\cE(q_1,\ldots,q_{n_1},p)\cdot
\cE(-p,q_{n_{1}+1},\ldots,q_{n_2})
=\cE(q_1,\ldots,q_{n_2}),
~~~~~~\text{for}~~~q_1+\cdots + q_{n_1}+p=0,
\label{filkeq1}
\eeqa
\beqa
\cE(q_1,\ldots,q_{n_1},p,q_{n_1+1},
\ldots,q_{n_2},-p)
=\cE(q_1,\ldots,q_{n_2}),
~~~~~~\text{for}~~~q_{n_1+1}+\cdots + q_{n_2}=0.
\label{filkeq2}
\eeqa
\end{subequations}

\end{appendix}

\begin{acknowledgments}
We are grateful to J.Zahn for helpful discussions.
This work is supported
in part by Korea Science and Engineering
Foundation Grant No.
R01-2004-000-10526-0,
and by the Korea Research Foundation
Grant funded by Korea Government(MOEHRD,
Basic Research Promotion Fund)(KRF-2005-075-C00009;
H.-C.K.)
\end{acknowledgments}
\vspace{3cm}
%\bibliographystyle{unsrt}
%\bibliography{bibli3}

\vspace{4cm}

\end{document}